\documentclass[letterpaper, 10 pt, conference]{ieeeconf}  

\IEEEoverridecommandlockouts                              

\usepackage{graphicx} 

\title{\LARGE \bf
Self-Similar Characteristics in Queue Length Dynamics: Insights from Adaptive Signalized Corridor
}

\author{Shakib Mustavee$^{1}$ and Shaurya Agarwal (Senior Member IEEE)$^{2}$
\thanks{*This work was not supported by any organization}
\thanks{$^{1}$Shakib Mustavee is a Ph.D. candidate with the Department of Civil, Environmental \& Construction Engineering, University of Central Florida, Orlando, Florida.
        {\tt\small shakib.mustavee@ucf.edu}}%
\thanks{$^{2}$Shaurya Agarwal is an Assistant Professor with the Department of Civil, Environmental \& Construction Engineering, University of Central Florida, Orlando, Florida.
        {\tt\small shaurya.agarwal@ucf.edu}}%
}

\begin{document}

\maketitle
\thispagestyle{empty}
\pagestyle{empty}

\begin{abstract}
Self-similarity, a fractal characteristic of traffic flow dynamics, is widely recognized in transportation engineering and physics. However, its practical application in real-world traffic scenarios remains limited. Conversely, the traffic flow dynamics at adaptive signalized intersections still need to be fully understood. This paper addresses this gap by analyzing the queue length time series from an adaptive signalized corridor and characterizing its self-similarity. The findings uncover a $1/f$ structure in the power spectrum of queue lengths, indicative of self-similarity. Furthermore, the paper estimates local scaling exponents $(\alpha)$, a measure of self-similarity computed via detrended fluctuation analysis (DFA), and identifies a positive correlation with congestion patterns. Additionally, the study examines the fractal dynamics of queue length through the evolution of scaling exponent. As a result, the paper offers new insights into the queue length dynamics of signalized intersections, which might help better understand the impact of adaptivity within the system.   


\end{abstract}

\section{Introduction}\label{sec:intro}

An adaptive system can dynamically modify its states through a regulatory process in response to the changing environment to maintain its optimum performance \cite{tyukin2011adaptation}. For example, intersections controlled by adaptive signal controllers finely tune the timing of red, yellow, and green lights to achieve specific operational objectives such as smooth traffic flow, throughput maximization, access equity, i.e., ensuring equity in green distribution to all approaches. The effectiveness of adaptive strategies is validated by various measures of effectiveness (MOEs) like throughput capacity, traffic delay, and queue lengths \cite{shafik2017field}. With the increasing availability of data and advancements in artificial intelligence (AI) technology, there is a growing trend in utilizing machine learning, deep reinforcement learning, and Q-learning for traffic signal control. However, despite the success of these black box models, a significant limitation remains in their interpretability. A crucial step in  AI-based optimization is designing a cost or reward function. In the current research, a typical cost or reward function for traffic signal control combines several MOEs weighted linearly \cite{zheng2019diagnosing}. This ad-hoc method presents various challenges, as determining the weight for each term is challenging, and even minor variations can lead to significantly different outcomes \cite{wei2019survey}. Moreover, data-driven AI approaches usually overlook transportation domain knowledge, limiting their robustness in real-world scenarios. Therefore, in addition to observing MOEs, it is critical to pay attention to the shift in the inherent characteristics of the dynamics when subjected to an adaptive control system.  




This raises the fundamental question: \textit{What are the inherent dynamical indicators that can be used to quantify the impact of adaptivity}? 
To address this question, the paper analyzes a corridor managed by intelligent traffic control units as a static network of dynamical systems. The study focuses on the statistical properties of its observables, with the queue length at each intersection being selected as the observable of interest. Managing queue length is crucial in achieving several operational objectives, making it a suitable observable to study.


\noindent \textbf{Why fractal dynamical approach?} 
Self-similar fractals often emerge as a collective behavior in systems with multiple interacting components, making them a hallmark of interconnected systems. Studying fractal properties can offer crucial insights into a system's stability, predictability, synchronization, and criticality. In traffic flow dynamics, adaptivity is characterized by synchronization, a collective behavior emerging from interactions among multiple subsystems \cite{arenas2008synchronization}. Achieving synchronization through diverse control mechanisms in traffic networks is crucial for ensuring stability and optimal performance \cite{pang2020coordinated}. At signalized intersections, various operational objectives, including smooth traffic flow, throughput maximization, and queue management, can be attained by ensuring the synchronization of traffic lights across neighboring road segments. Therefore, investigating how self-similar behaviors in the traffic flow dynamics of signalized intersections change might provide critical insights into their signal coordination, i.e., adaptivity.

\noindent \textbf{Contributions:} The contributions of this article are diverse and multifaceted. These contributions are detailed as follows:

\begin{itemize}
        \item The study characterizes the spectrum of queue length time series and finds pink noise, a self-similar fractal characteristic across all the intersections of the corridor.
        \item The study computes scaling exponents from the queue length time series to quantify self-similarity,
        \item The study unveils periodic trends in scaling exponent and its positive correlation with congestion levels computed from queue lengths. 
        \item The study finds that time-varying patterns of local scaling exponents of adjacent intersections have similar patterns regardless of size.
        \item The study unveils the evolution of self-similarity in queue length dynamics and finds its relationship with congestion. This finding holds potential as a valuable tool for quantifying the adaptivity of signal controllers.
\end{itemize}

 The structure of the paper is as follows. Section~\ref{sec:litrev} discusses the relevant literature on fractal analysis of traffic networks. Section~\ref{sec:MathData} provides background on mathematical tools, Section~\ref{sec:res} provides details of the case study and results, Section~\ref{sec:disc} provides a rich discussion on obtained results, and Section~\ref{sec:conc} provides concluding remarks.

\section{Literature Review and Knowledge Gaps} \label{sec:litrev}
\noindent \textbf{Indicators of adaptivity:} Statistical properties like scaling exponent, Lyapunov exponent, and entropy of time series are often investigated to quantify adaptivity across diverse domains. For instance, ensemble maximum Lyapunov exponent and alignment index were used as indicators of adaptivity for adaptable Hamiltonian neural networks \cite{han2021adaptable}. In cognitive science, multi-scale entropy analysis (MSE) and detrended fluctuation analysis (DFA) are employed to measure adaptability in tasks like jump landing and dynamic postural stability under varying conditions \cite{johnson2022relationship}. Adaptability in human gait is assessed by the DFA scaling exponent of human inter-stride intervals time series in \cite{vaz2019synchronization}. Multifractal indices were used to verify the adaptability of the online classification model of internet traffic flows \cite{tang2023online}. For the stability and adaptivity analysis of wind power generators integrated into large interconnected electric grids, a dynamic equivalent approach is introduced in \cite{zhou2018robustness}. Adaptability in cooperative autonomous driving, particularly in competitive scenarios such as highway merging and exiting ramps within mixed-traffic environments, is measured by domain adaptation matrix \cite{valiente2022robustness}. 

\noindent \textbf{Fractal analysis of traffic flow dynamics:} Fractal behavior in traffic dynamics has been studied for a long time. Notably, the initial hints of fractal behavior in traffic dynamics surfaced almost five decades ago in $1978$, when T. Musha and E. Higuich discovered that, similar to various granular flow dynamics, velocity-velocity correlation functions within expressway traffic in Japan exhibited a self-similar $1/f$ low-frequency behavior. However, it has not been fully explored in the field of transportation engineering, as noted in \cite{laval2023self}. The existing fractal literature on traffic flow dynamics can be broadly divided into two distinct perspectives: (i) physics-inspired and (ii) statistical. Within the former perspective, self-organized criticality (SOC), a prominent paradigm of fractal analysis, is often used to analyze the underlying physics of traffic systems and involves identifying power law relationships across various traffic networks, especially concerning traffic congestion scenarios. The SOC theory for traffic analysis was first introduced in 1996 by Paczuski and colleagues \cite{paczuski1996self}. They conducted an analytical study of phantom traffic jams that develop due to intermittent stop-and-go traffic patterns and derived power law characteristics in the spectrum of these traffic jams. While the SOC theory aims to derive fractal characteristics from analytical traffic flow models, a new perspective emerged in the early 2000s, focusing on the statistical approach. Researchers used statistical analysis tools like DFA, Multifractal Detrended Fluctuation Analysis (MFDFA), Multifractal Detrended Cross-correlation Analysis (MFDXA), Rescaled Range (R/S) analysis, autocorrelation function (ACF), and Hölder exponent to examine various traffic time series and characterize fractal behavior. These tools are applied to various traffic time series data, including speed, flow, and density, to reveal correlations between traffic variables and identify long-range dependence. While the SOC-based fractal traffic analysis primarily aims to identify and characterize phenomena like congestion, meta-stability, and state transitions, the time series fractal analyses focus on revealing long and short-range correlations, memory effects, and persistence. Fractal analysis has shown that traffic arrival patterns on highways under moderate to heavy traffic conditions exhibit self-similar behavior \cite{meng2009self}. Therefore, a Poisson distribution may not be suitable for modeling vehicle arrival patterns in such conditions. Additionally, previous studies have identified that vehicle arrival patterns in an isolated signalized intersection demonstrate self-similar behavior \cite{khoo2019study}. However, there is a gap in the literature when it comes to characterizing the fractal behavior of a signalized corridor consisting of consecutive adaptive signalized intersections. The current study aims to fill this gap.

\section{Concepts and Mathematical Background} \label{sec:MathData}
This section details essential concepts and mathematical tools crucial for interpreting the study. 
The discussion encompasses key concepts such as fractal behavior and scale invariance, followed by an introduction to mathematical tools, including power spectral density and DFA. 

 
\noindent \textbf{What is fractal?} Fractals are complex geometric structures that display self-similarity, meaning that they contain repeating patterns or smaller copies of themselves at different scales \cite{mandelbrot1982fractal}. 

\noindent \textbf{Fractal processes:} Fractals are associated with geometric shapes and objects, but many processes also exhibit self-similar features when analyzed over time. A self-similar time series displays the same statistical properties across different time scales, much like geometric self-similarity. A self-similar time series $X(t)$ satisfies the following condition: $X(ct) = c^{\alpha}X(t)$, where $\alpha$ is a scaling exponent that indicates a scale-invariant nature of the time series.


\noindent \textbf{What does the scaling exponent signify?}
The scaling exponent can provide valuable insights into a process, such as its long-range dependence, persistence, and stationarity. In certain systems, self-similarity is believed to be an emergent behavior resulting from interactions between multiple components. Consequently, the scaling exponent can also aid in comprehending the coordination, synchronization, stability, and adaptability of interconnected systems.

\noindent \textbf{How to measure the scaling exponent?}
Various mathematical techniques have been employed to characterize the fractal behavior of time series. These techniques can be classified based on the nature of analysis as follows: (i) Time domain methods, including rescaled range analysis (R/S), autocorrelation analysis (AC), scaled windowed variance analyses (SWV), dispersion analysis, and detrended fluctuation analysis (DFA); (ii) Frequency domain methods, such as power spectral density (PSD) and coarse-graining spectral Analysis (CGSA); and (iii) Time-frequency domain analysis techniques, like short-time Fourier Transform (STFT) and fractal wavelet analysis \cite{eke2002fractal}. This study utilizes both frequency and time domain methods, specifically PSD and DFA, to characterize the fractal behavior of queue length time series. An introduction to PSD and DFA follows below.

\subsubsection{\textbf{Power Spectral Density (PSD)}} \label{sec:PSD}
PSD is a useful tool for understanding how a signal's power is distributed across different frequencies. While various methods exist for computing PSD, this study employs the Fast Fourier Transform (FFT) to calculate it accurately. The signal's length (\(N\)) is determined initially. Subsequently, the FFT of the signal (\(x(t)\)) is computed, resulting in \(X(f)\). If \({fs}\) is the sampling frequency, the power spectrum \(S(f)\) as a function of frequency (\(f\)) is then calculated using the formula:

     \[S(f) = \frac{1}{{fs} \cdot N} \cdot |X(f)|^2 \]    

The power spectrum $S(f)$ of a fractal time series exhibits scale invariance and follows a power law,

\[ S(f) \sim f^{-\beta} \]

The parameter \(\beta\), known as the spectral density exponent, plays a crucial role in characterizing the behavior of time series data. The power law translates into a straight line in a log-log plot, with \(\beta\) serving as the slope. This exponent quantifies the level of correlation present within the time series, with white noise (\(\beta = 0\)) indicating no correlation and Brownian noise (\(\beta = 2\)) reflecting a strong correlation. Furthermore, pink noise (\(\beta = 1\)), also known as \(1/f\) noise, represents a delicate balance between white noise and Brownian noise, signifying equal energy per octave of frequency.

\subsubsection{\textbf{Detrended Fluctuation Analysis (DFA)}} \label{sec:DFA_math}

The DFA exponent $(\alpha)$ measures the scaling exponent serving as a metric for self-similarity in nonstationary time series. DFA calculates the fluctuation function as the RMS variance from the local trends at different time scales, and the slope of the fluctuation function and scale in a log-log plot represents the scaling exponent $(\alpha)$.

\[
F(s) = \left[ \frac{1}{N_s} \sum_{v=1}^{N_n} F^2(n, v) \right]^{1/2}.
\]

In the presence of power-law correlation, \(F(s) \propto s^{\alpha}\), where $\alpha$ is the estimation of the scaling exponent. Therefore, \( \alpha \) can be computed by extracting the slope of the regression line through a log-log plot of \(F(s)\) against the scale \(s\). DFA exponent $\alpha$ provides insight into correlation structure and persistence in time series. In addition, several authors have empirically shown the connection between the DFA scaling exponent and the power exponent of anomalous diffusion. The relationship helps characterize the time series as Fractional Brownian motion (fBm) or Fractional Gaussian noise (fGn).

\noindent \textbf{Relation between \(\alpha\) and \(\beta\):} DFA exponent \(\alpha\) characterizes power law correlation in the time domain while \(\beta\) characterizes power law correlation in the frequency domain on the same signal. The relation between the two parameters is determined by the Wiener-Khinchin theorem, which states the power spectrum of a stationary stochastic process is analogous to the Fourier transform of the corresponding autocorrelation function. Therefore, \(\alpha\) is linearly related to the exponent \(\beta\) as follows: \(\beta\) = \(2\alpha-1\).

\section{Case Study} \label{sec:res}

The fractal analysis is performed in \textbf{three phases}, each playing a pivotal role in reaching the conclusions. The first phase analysis involves characterizing the power laws within queue length time series at signalized intersections, employing PSD analysis. The second phase calculates the local scaling exponent of these time series. The third phase conducts a time-varying analysis of the local scaling exponent in the third phase, strengthening the insights gained from the second phase. Before diving deep into results and discussion, let's look at the data used in the case study.

\subsection{\textbf{Data Description}} \label{subsec:data}

\begin{figure}[!htbp]\center
\includegraphics[width=1.0\linewidth]{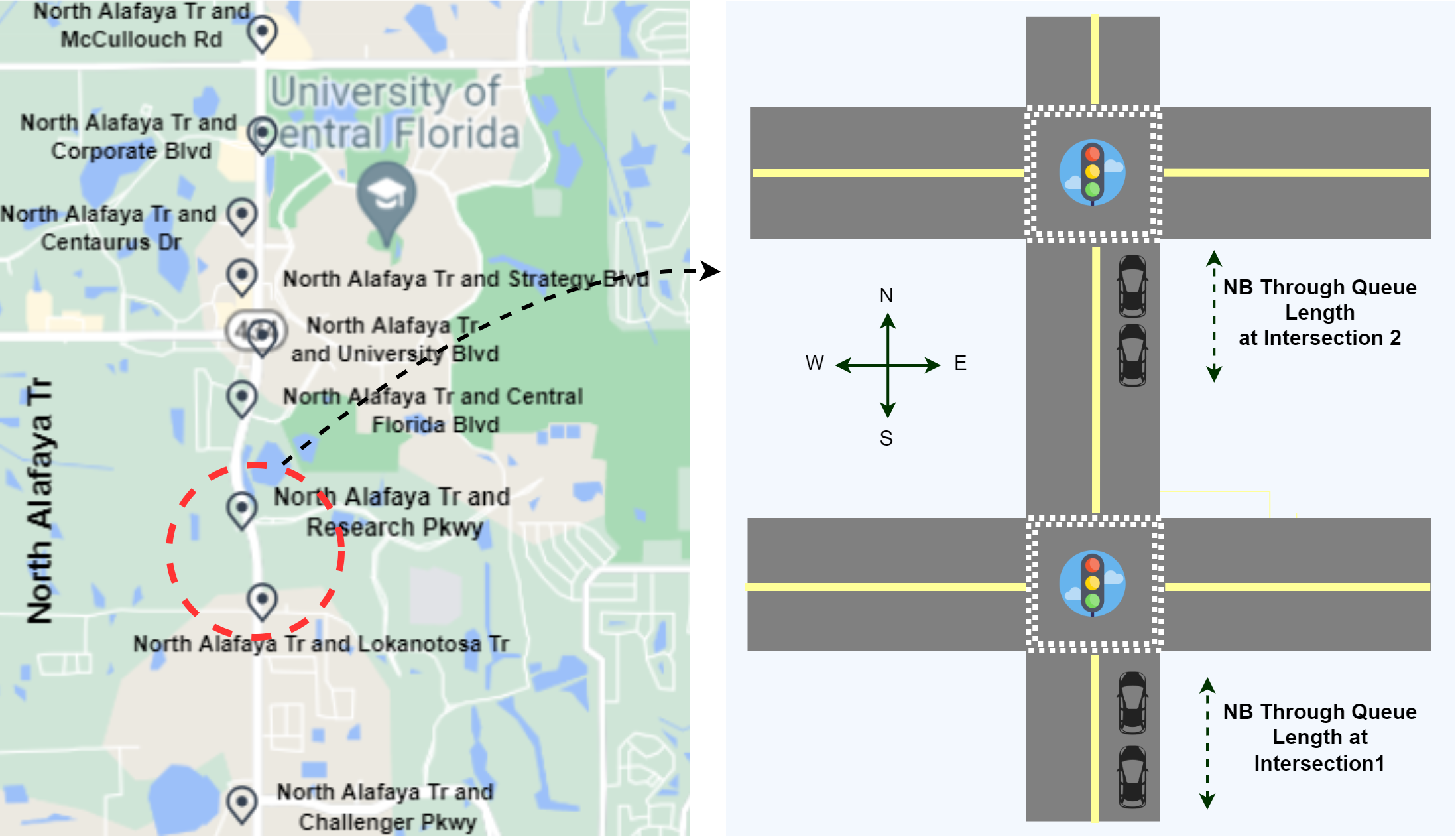}
\caption{Location of intersections on Alafaya corridor (left) and illustration of the corridor as a dynamical system (right). Northbound (NB) through queue length at each intersection is measured in distance units.}
\vspace{-4mm}
\label{fig:corridor}
\end{figure}

The dataset comprises queue lengths of $9$ adaptive traffic signals managed by \textit{InSync} recorded between December 18, 2017, and February 14, 2018, at the Alafaya Trail (SR-434) corridor in East Orlando, FL, shown in Figure~\ref{fig:corridor}. This study primarily investigates major roads, i.e., north through movements (NT), as previous research has shown the effectiveness of adaptive signals for these routes \cite{shafik2017field}. The queue length time series is generated by recording the maximum queue length every two minutes, which approximates the cycle length. For a visualization of the time series, refer to \cite{das2023koopman}. 
The study followed the outlier mitigation and data noise reduction methods suggested in \cite{rahman2021real} and \cite{shabab2023dynamic}.

\subsection{\textbf{Power Law Scaling Characterization}}
The first analysis phase characterizes the power law scaling properties within queue length dynamics. It estimates the spectral exponent $\beta$ that provides valuable insights into the self-similarity of the queue length time series. 

\subsubsection{\textbf{PSD Analysis and Addressing Sensor Failure}}
\begin{figure}
        \includegraphics[width=\linewidth]{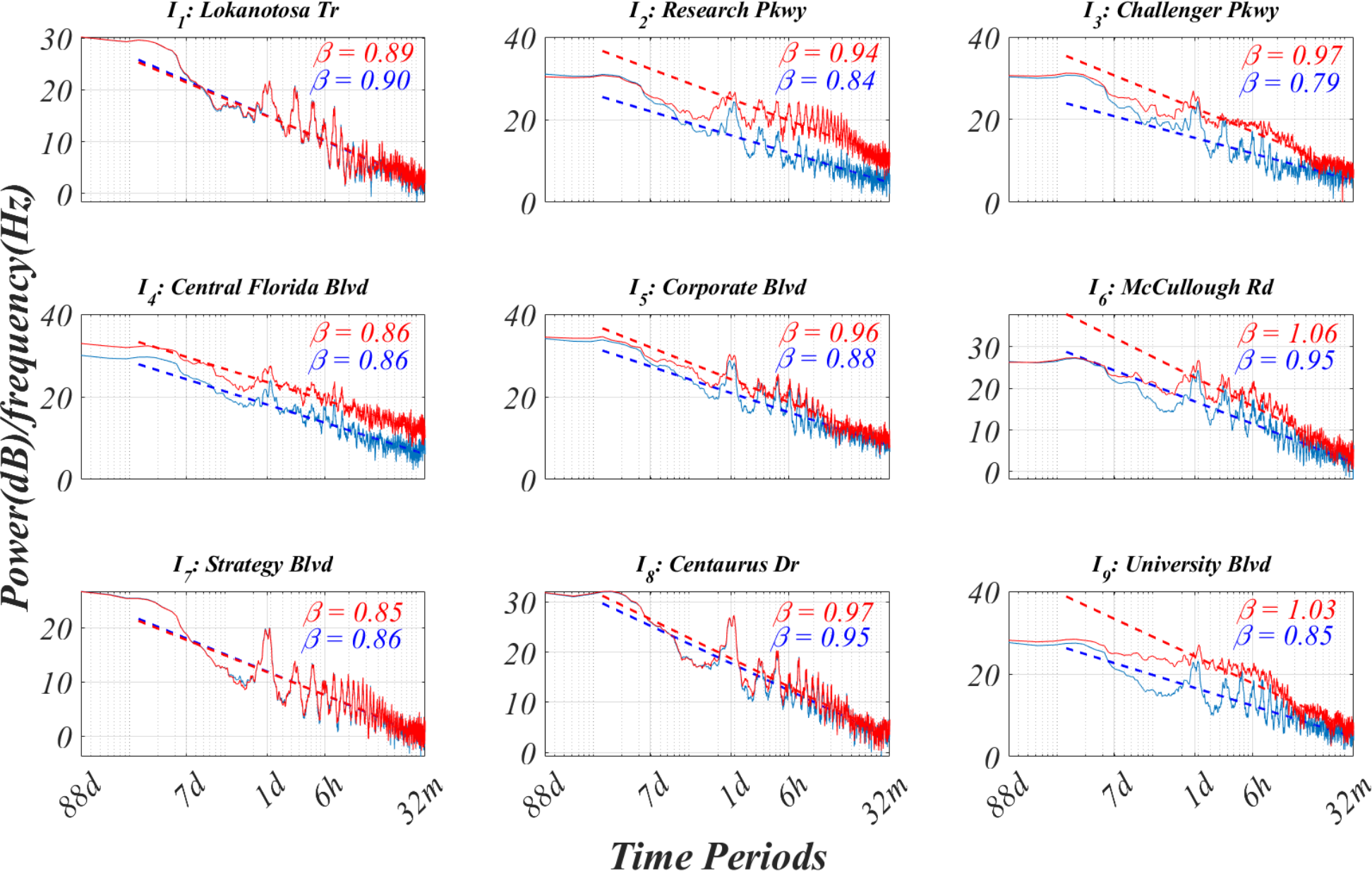} 
        \caption{Illustrates the power spectrum computed from the raw data (red) and outlier removed data (blue).}
        \label{fig:Regression}
    \end{figure}

Figure~\ref{fig:Regression} illustrates the power spectrum of queue lengths time series recorded at signalized intersections. The data is aggregated at a cycle level of $2$ minutes and spans $88$ days. We have noted that the power spectrum of any queue length time series over a week exhibits a $1/f$ structure. However, the study uses the longest available dataset to ensure the reliability of the analysis. The figure also provides a comparative analysis between two scenarios: (i) PSDs estimated from raw data (in yellow) and (ii) PSDs estimated from outliers removed data (in blue). The comparison demonstrates how spectral analysis might be used to evaluate the impact of data misreadings. The figure shows power spectrum changes after outlier removal. The power spectra of Lokanotosa Tr, Strategy Blvd, and Centaurus Dr superimpose for both cases, while the spectra of Research Pkwy, McCullough Rd, and University Blvd deviate from a $1/f$ structure. Additionally, the queue length power spectra of University Blvd and Research Pkwy estimated from raw data exhibit two distinct regimes of power exponents: one from 14 days to 6 hours and the other from 6 to 2 hours. The emergence of these two separate power-law scales may be attributed to the influence of false readings.



    \begin{figure}
        \includegraphics[width=\linewidth]{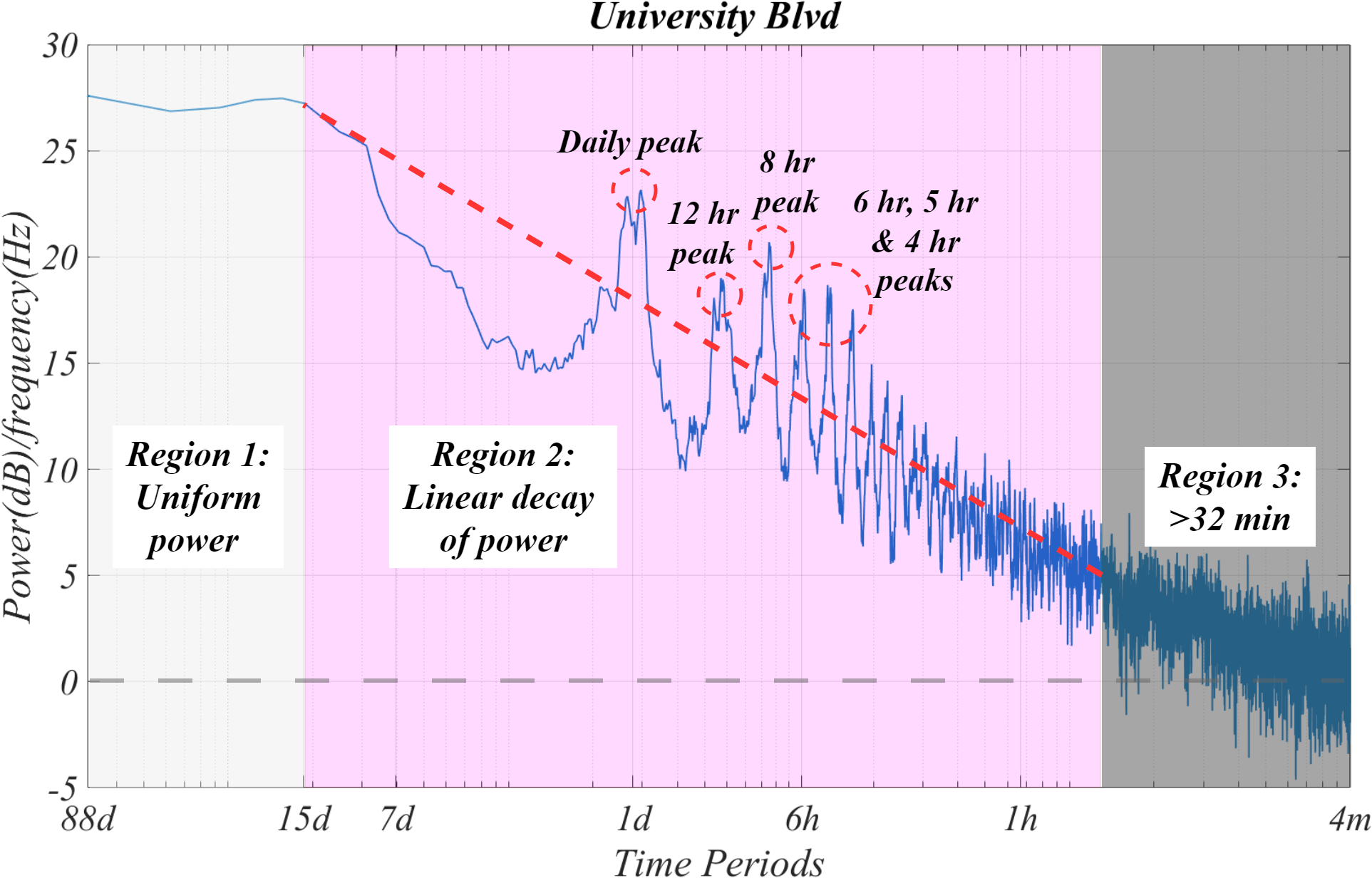} 
        \caption{Queue length power spectrum of the University Blvd intersection is shown here as an example for examining distinct spectrum regions.} 
        \label{fig:zoomedPSD}
    \end{figure}

We analyze various regions of the spectrum as depicted in Figure~\ref{fig:zoomedPSD} by looking closer at the PSD of University Blvd. The entire plot can be segmented into three distinctive regions: (i) \textbf{Low-frequency region:} The width of the low-frequency region varies depending on the number of days analyzed. Examining various time series lengths, we determined that the first six harmonics consistently fall within this low-frequency region,  (ii) \textbf{Linear decay region:} A $1/f$ profile, showing a decreasing trend within the frequencies corresponding to $14$ days to $30$ minutes. The sharp spikes represent system frequency for signalized intersections such as $24$ hr, $12$ hr, $8$ hr, $6$ hr, and $4$ hr as reported in \cite{das2023koopman} (iii) \textbf{High-frequency region:} Frequencies higher than $32$ minutes fall in this region. The region is excluded while fitting the line.

\subsubsection{\textbf{Extracting Spectral Exponent}}
The slopes of fitted lines are denoted by $\beta$. In the line fitting process, we discarded both low and high-frequency regions. The low-frequency region, up to 14 days, indicates white noise. Thus, periods ranging from $88$ to $14$ days were excluded while fitting the line. Removing high frequencies is a common practice in the literature \cite{eke2002fractal}. Consequently, periods higher than 32 minutes were omitted during line fitting. We show $\beta$ in Figure~\ref{fig:Regression}. Therefore, the straight lines are fitted between $14$-day to the $32$-minute, i.e., linear decay region.

\subsection{\textbf{Interpreting DFA Scaling Exponents Patterns}} \label{subsub:Result_DFA}
The second phase uses DFA to unravel the intricate relationship between congestion patterns and local scaling exponents within queue length dynamics. 

\subsubsection{\textbf{Verifying Scaling Exponent}} \label{sec:DFA}
Now, we verify the spectral exponent $(\beta)$ using its theoretical relationship with DFA exponent $(\alpha)$. The study utilizes the same dataset. The DFA method uses the power of 2 scale lengths, and usually, the length of the time series is chosen $4$ to $8$ times the maximum scale. We choose maximum scale length as $2^{13}$ samples and time series length $2^{15}$, i.e., 45.5 days. For calculating $\alpha$, we use a time series that spans 46 days from January 1, 2018, to February 16, 2018. Table~\ref{tab:compare} compares spectral exponent ($\beta$) and DFA exponents($\alpha$). Here, $\tilde{\beta}= 2\alpha -1$ is computed using the theoretical relationship for comparing $\alpha$ and $\beta$. The table shows that the difference between $\beta$ and $\tilde{\beta}$ was less than $5\%$ at $7$ out of $9$ intersections, confirming the theoretical relationship from the Wiener-Khinchin theorem. While a longer time series analysis provides an overall understanding of the fractal dynamics, it does not capture the changes in the local fractal behavior of a time series. Since traffic situations are continuously evolving, practical applications often require knowledge of instantaneous changes in system behavior. Therefore, we use DFA to study the local fractal behavior by examining how daily scaling exponent changes with congestion. 

    
\begin{table}
\caption{Comparison of $\beta$ and $\tilde{\beta} =2\alpha-1$}
    \centering
    \begin{tabular}{|c c c| c c c|}
    \hline
     \textbf{Intersection} & $\beta$ & $\tilde{\beta}$ & \textbf{Intersection}  & $\beta$ & $\tilde{\beta}$\\
     \hline
     Lokanotosa  & 0.90 & 0.92 & Research & 0.84  & 0.84 \\
     Challenger & 0.79 & 0.88 & Central FL & 0.86 & 0.76\\
      Corporate  & 0.88 & 0.86 & McCullough & 0.95 & 0.95\\
      Strategy & 0.86  & 0.82 & Centaurus & 0.95 & 0.94 \\
      University & 0.85 & 0.82 & & &\\
      \hline
\end{tabular}
    \label{tab:compare}
\end{table}

\subsubsection{\textbf{Estimating Local Scaling Exponent}} \label{sec:DFA_local}
To understand local fractal characteristics, We estimate the local scaling exponent $(\alpha)$ by analyzing small segments of the entire time series. We chose the maximum scale length of $2^{8}=256$ with a time series length of $4 \times 256 = 1024$ (approx. 34 hours). We selected the $34$-hour window from $7$ AM to $5$ PM the following day. The window aims to encompass the maximum number of morning and evening peaks. We define the local scaling exponent calculated within a $34$-hour window as the daily scaling exponent. 


\subsubsection{\textbf{Correlation between Scaling Exponent and Congestion}}

\begin{figure}
        \includegraphics[width=\linewidth]{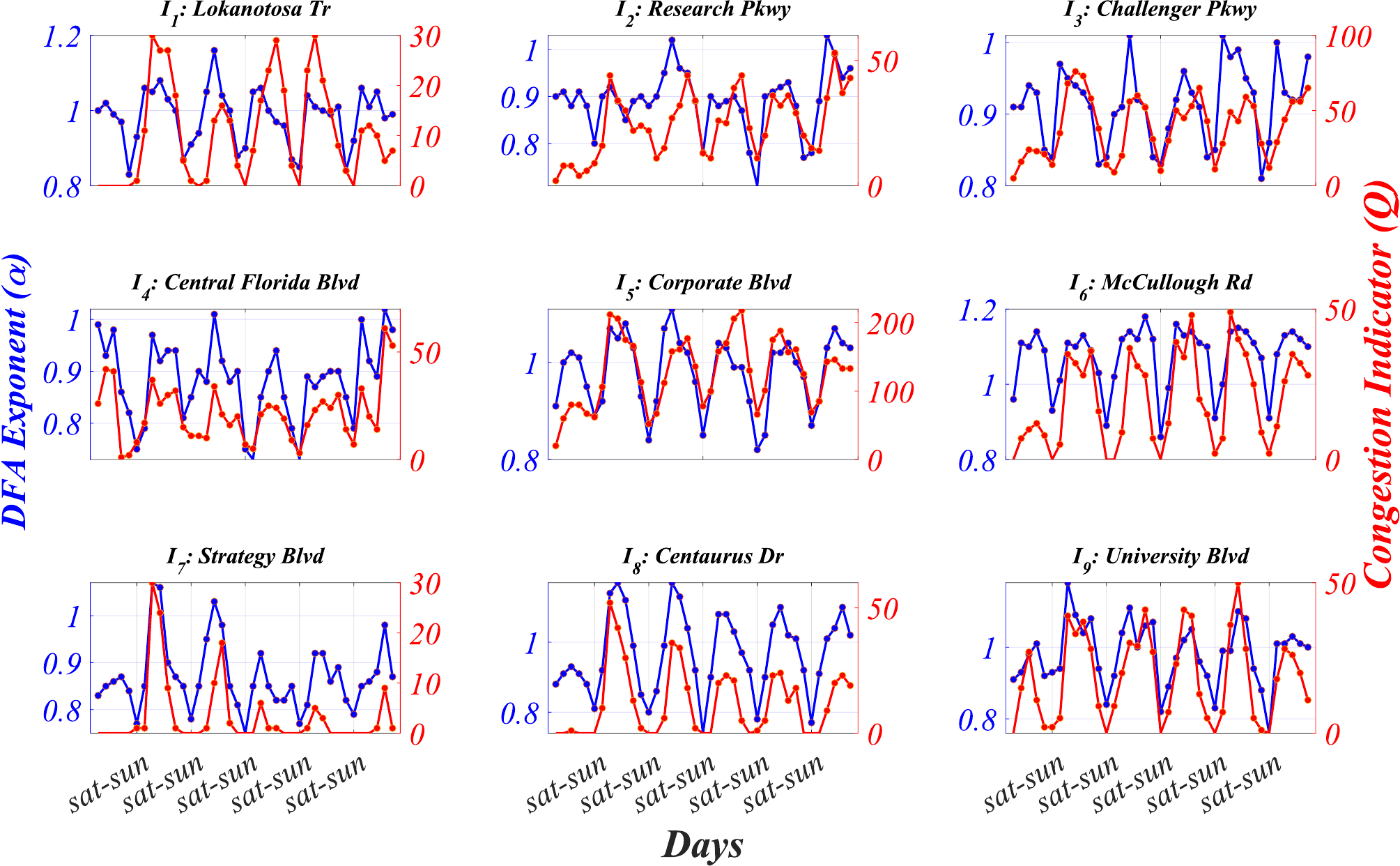} 
        \caption{Evolution of daily congestion and scaling exponent.}
        \label{fig:more25}
    \end{figure}
\begin{figure}
        \includegraphics[width=\linewidth]{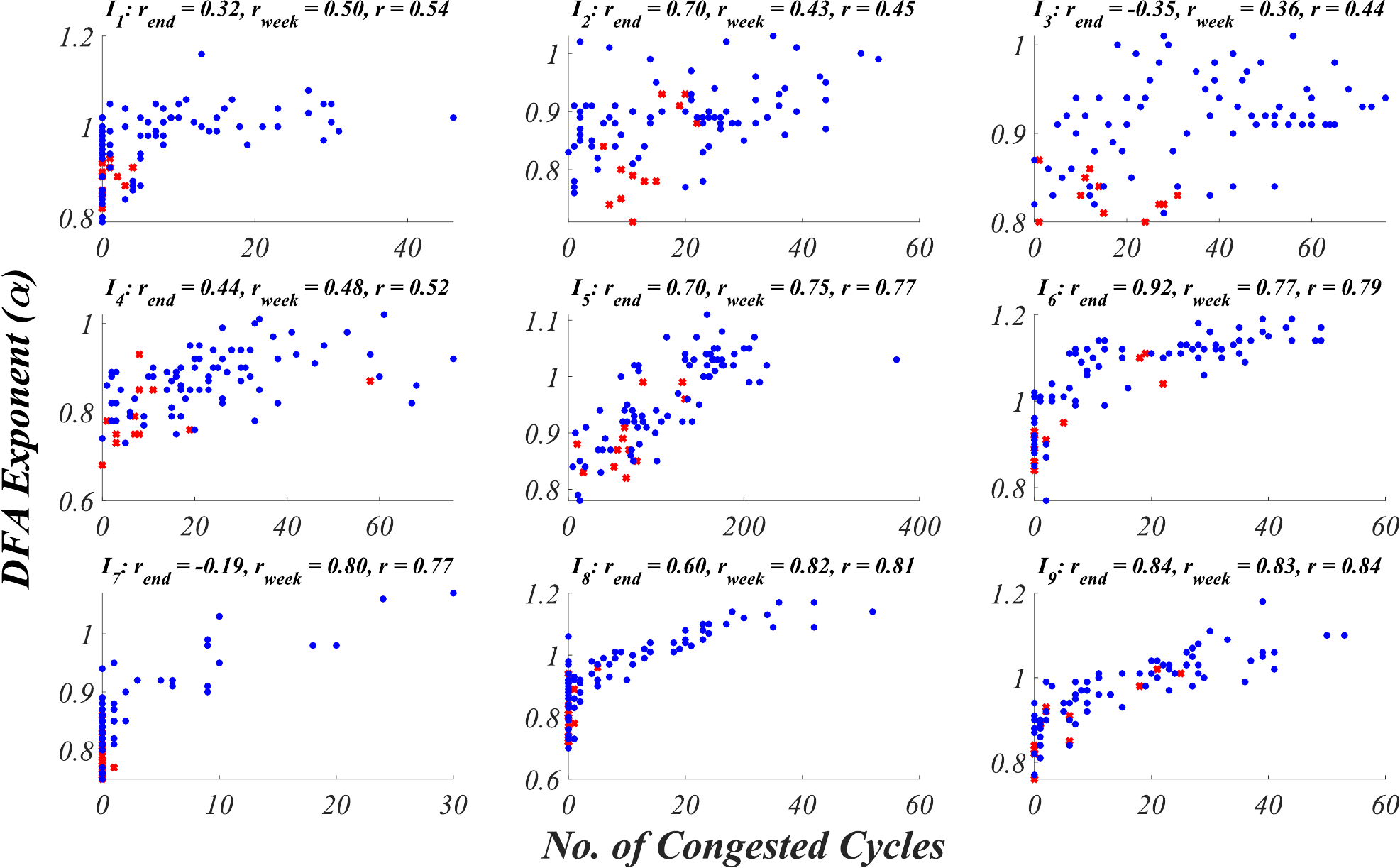}
        \caption{Correlation between $Q$ and $\alpha$. The red crosses represent weekends, and the blue dots represent weekdays. Pearson correlation coefficient between $Q$ and $\alpha$ during weekends, weekdays, and overall are $r_{end}$,  $r_{week}$ and $r$ respectfully.}
        \label{fig:corr}
    \end{figure}

Each intersection has a maximum queue length it can handle during the red signal phase, which is known as its capacity. Congestion is defined as the queue length reaching a certain percentage of the maximum limit. Here, the congestion indicator ($Q$) is defined by the number of cycles within an observation window exceeding the threshold. We consider the threshold as $60\%$ of the total capacity ($40$ vehicles), i.e., $25$ vehicle. Figure~\ref{fig:more25} shows how congestion indicator $(Q)$ and daily scaling exponent ($\alpha$) evolve. Both the parameters display similar weekly-weekend patterns. To investigate the relation $Q$ between $\alpha$, we estimate their correlation. The queue length data spans 88 days from January $17$, $2017$, to March $20$, $2018$. Consequently, for each intersection, there are $88$ pairs of DFA exponent ($\alpha$) and congestion indicator ($Q$) computed. Each pair of $\alpha$ and $Q$ are plotted as a scatter plot shown in Figure~\ref{fig:corr}. A positive correlation exists between $(\alpha)$ and $(Q)$ across all intersections. However, the correlation weakens on weekends and is negative at a few intersections.




\subsection{\textbf{Time-Dependent Scaling Exponent $\alpha(t)$}}
The third analysis phase observes the evolution of the scaling exponent through $(\alpha (t))$.

\subsubsection{\textbf{How to Estimate $\alpha(t)$}}
\begin{figure}
        \includegraphics[width=\linewidth]{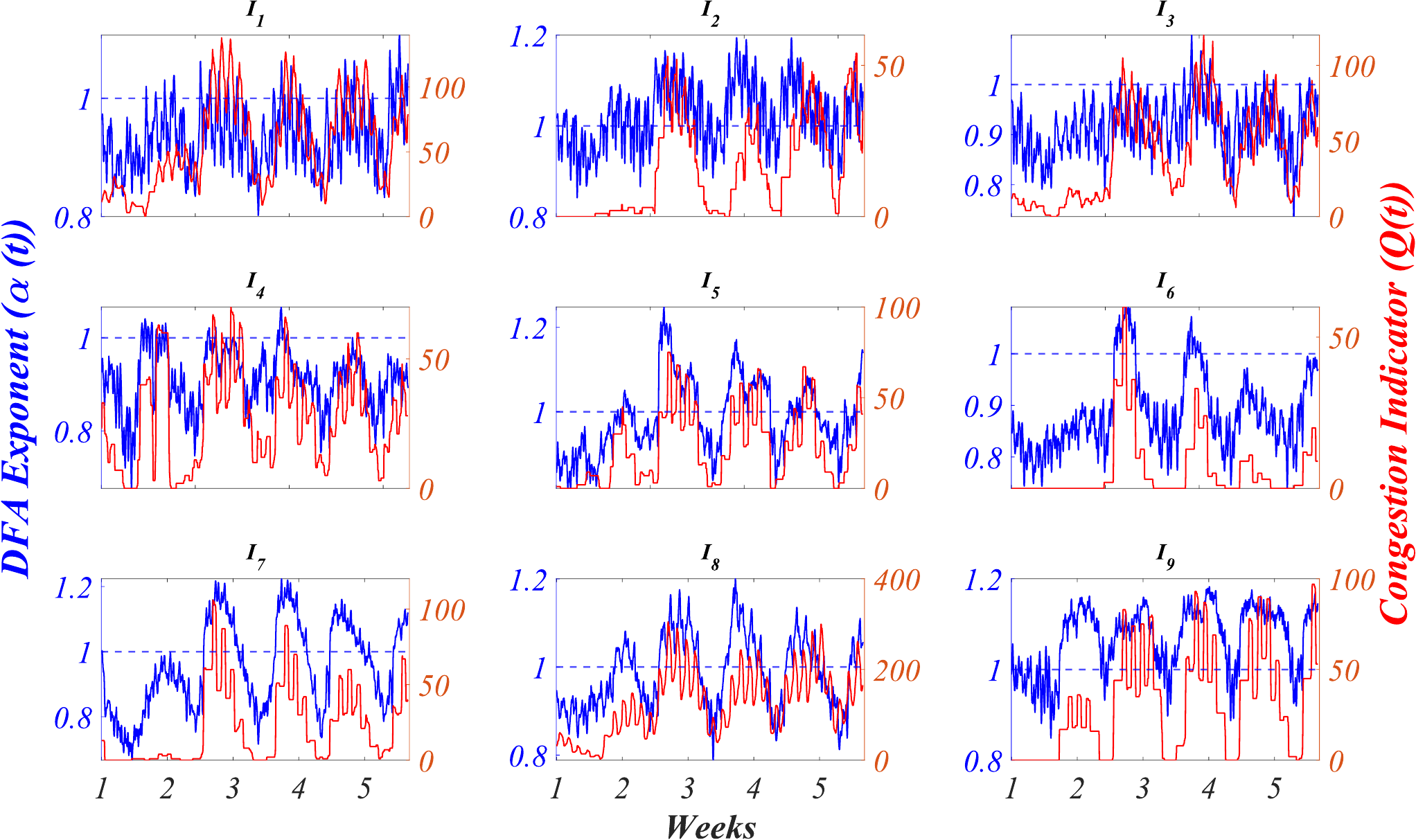}
        \caption{Illustrates the evolution of the local scaling exponent and congestion patterns over a month, highlighting the similarity in profiles among neighboring intersections.}
        \label{fig:H(t)}
    \end{figure}

The analysis keeps the window size the same as before, i.e., $34$ hours ($1024$ samples). However, it changes the starting and ending points of the window at each iteration. The time-dependent analysis of the scaling exponent shifts the window by $15$ samples, i.e., $30$ minutes at each iteration. Here, the threshold for $Q$ is assumed $50\%$ of the capacity. Figure~\ref{fig:H(t)} shows the time series of the time-dependent scaling exponent across the corridor. The figure helps us understand the dynamic relationship between $Q$ and the local scaling exponent. The fluctuations in $\alpha(t)$ can be explained by changes in the congestion indicator. The increase in the number of congested cycles within a window follows an increase in $\alpha(t)$. The figure also shows the evolution of $Q$ in different intersections.



\subsubsection{\textbf{Characterizing Intersections Based on Dynamic Scaling Exponent}}
Based on the pattern of $Q(t)$ and $\alpha(t)$ we identify three distinct types of dynamic behaviors:

(i) High daily fluctuation: Sharp fluctuations in $\alpha(t)$ are observed at Locanotosa Rd, Research Pkwy, and Challenger Pkwy, which might be related to the significant increase in traffic demand during peak hours marked by spikes-like patterns in $Q$ during weekdays. 

(ii) Plateau trends: At Central FL Blvd, Strategy Blvd, and University Blvd, the local trend remains flat or plateau-like due to consistent demands during weekdays.

(iii) Plateau trends within the Brownian regime: During weekdays, there are flat or plateau-like local trends observed at the intersections of Corporate Blvd, Centaurus Dr, and McCullough Rd. These intersections exhibit more consistent values of $\alpha(t)$ throughout the weekdays than the second category, with $\alpha(t)$ oscillating around 1.1 to 1.2. This indicates the presence of queue length dynamics operating in the Brownian regime.

\section{\textbf{Takeaways and Discussion}} \label{sec:disc}
The first phase of analysis reveals the fractal nature by analyzing the long-term behavior of queue length dynamics. The second phase investigates the local fractal behavior through scaling exponent $(\alpha)$. Within the current literature, for traffic flow time series $0.5<\alpha<1$ is interpreted as synchronized flow, while an $\alpha>1$ indicates traffic jams, and several studies have shown that a decrease in demand decreases $\alpha$ \cite{krause2017importance}, \cite{peng2010long}. Our study reveals that on weekends, $\alpha$ drops to around $0.7-0.8$, while during peak hours of weekdays, $\alpha$ rises beyond $1.0$, which is consistent with existing literature. Moreover, the study finds a positive correlation between $\alpha$ and congestion indicator $(Q)$, suggesting adaptivity affects $\alpha$. Queue length depends on the number of vehicles arriving at the red light and departing at the green light. Adaptive intersections at a corridor use real-time vehicle arrival data to adjust green light times, which is also influenced by the outflow of the previous intersection, controlled by an adaptive controller. While queue length in fixed-time signals depends solely on traffic demand, in adaptive intersections, it depends on the efficiency of the adaptive mechanism in addition to the traffic demand. Therefore, we hypothesize that the evolution of $\alpha$ might be able to track adaptivity due to its positive correlation with $Q$ computed from queue length. Therefore, the third phase estimates the evolution and finds $\alpha(t)$ and $Q(t)$ follow similar increasing and decreasing patterns as shown in Figure~\ref{fig:H(t)}. Furthermore, the intersections are classified empirically into three categories based on their evolutionary pattern, each comprising three consecutive intersections. We plan to utilize time series classification tools to support this empirical classification in future research. Consecutive intersections showing similar $\alpha(t)$ patterns, despite differences in size and demand, may indicate synchronization along the adaptive corridor.



\section{Conclusion} \label{sec:conc}
This study investigates the self-similar fractal behavior of the queue length and finds a positive correlation between the scaling exponent of a region and its congestion indicators. Additionally, there is a similarity in the $\alpha(t)$ patterns among neighboring intersections. This information can be valuable for analyzing, monitoring, and evaluating the system's performance. The variable nature of the local scaling exponent indicates the multifractal nature of the time series which might reveal more insights into the adaptivity of the system. Therefore, further research is necessary to characterize the multifractal behavior of queue length dynamics. Additionally, the self-similar behavior of other system observables, such as arrival patterns and green time ratios, should be explored in future studies.
\bibliographystyle{IEEEtran}
\bibliography{fracRef}

\end{document}